\documentstyle[floats,prd,aps,eqsecnum,12pt]{revtex}

\makeatletter
\newbox\tempboxa
\newdimen\captionboxsubcount
\def\capsize#1{\captionboxsubcount=#1pt}
\newdimen\captionboxsub
\captionboxsub=\hsize \advance\captionboxsub by
-\captionboxsubcount \advance\captionboxsub by
-\captionboxsubcount \long
\def\@makecaption#1#2{
\setbox\@tempboxa\hbox{#1 #2} \ifdim \wd\@tempboxa >\captionboxsub
\rightskip=\captionboxsubcount \leftskip=\captionboxsubcount #1 #2
\else \hbox to\hsize{\hfil\box\@tempboxa\hfil} \fi}
\makeatother
\capsize{30}

\begin{document}

\begin{titlepage}
\begin{flushright}
\begin{minipage}{5cm}
\begin{flushleft}
\small
\baselineskip = 13pt YCTP-P1-00
\end{flushleft}
\end{minipage}
\end{flushright}

\begin{center}
\Large\bf Enhanced Global Symmetry Constraints on $\epsilon$ Terms
\end{center}
\vfill \footnotesep = 12pt
\begin{center}
\large
 Zhiyong {\sc Duan} \footnote{Electronic address:
 {\tt zhiyong.duan@yale.edu}} \quad P.~S. {\sc Rodrigues da Silva}
 \footnote{Electronic address:
{\tt psr7@pantheon.yale.edu}} \quad
 Francesco {\sc Sannino} \footnote{ Electronic address
: {\tt francesco.sannino@yale.edu}}
\\~\\
{\it
Department~of~Physics,~Yale~University,~New~Haven,~CT~06520-8120,~USA.}
\end{center}
\vfill
\begin{center}
\bf Abstract
\end{center}
\begin{abstract}
\baselineskip = 17pt Recently it has been proposed that the physical
spectrum of a vector-like gauge field theory may exhibit an
enhanced global symmetry near a chiral/conformal phase transition
\cite{ARS}. The new symmetry is related to the possibility,
supported by various investigations, that a parity-doubled spectrum
develops as the number of fermions $N_f$ is increased to a critical
value above which it is expected that  the symmetric phase is
restored.
 We show that parity-doubling together with the associated
enhanced global symmetry severely constrains the $\epsilon$ terms
of the effective Lagrangian involving Goldstone bosons as well as
massive spin-1 particles. We extend our analysis to underlying
fermions in pseudo-real representations of the gauge group.
\end{abstract}
\begin{flushleft}
\end{flushleft}
\vfill
\end{titlepage}

\section{Introduction}

In recent years much progress has been achieved in understanding some
aspects of strongly interacting gauge theories. In particular the phase
structure of supersymmetric gauge theories as function of the number of
matter flavors has been analyzed using effective Lagrangian approaches,
duality arguments and a variety of consistency checks \cite{IS}.

{}For ordinary gauge theories the situation is much less firm and further
investigations are required to clarify the strong dynamics. Recently the
phase structure for ordinary gauge theories has been investigated in Ref.~{\
\cite{AS,ATW,SS}}. The infrared behavior of the underlying theory is
expected to be very different according to the number of massless matter
fields $N_{f}$ coupled to the gauge fields. A reasonable scenario is that
for low values of number of flavors the theory confines and chiral symmetry
breaking occurs. On the other side for large number of flavors the theory
loses asymptotic freedom. It is expected that in between these two regimes
there is a {\it conformal window} \cite{ATW}. In this region the theory does
not confine, chiral symmetry is restored, and the theory acquires a long
range conformal symmetry. While the upper limit of the conformal window is
uniquely fixed by identifying it with the point above which asymptotic
freedom is lost, there is still no complete agreement on the lower one. It
has been argued that the transition from a confining (possibly with broken
chiral symmetry) to a conformal theory in a $SU(N)$ gauge theory, as a
function of the number of flavors $N_{f}$, takes place for $N_{f}\approx 4N$
\cite{ATW,SS,ACS}. However recent lattice simulations \cite{mawhinney} seem
to indicate that the amount of chiral symmetry breaking decreases
substantially (for $N=3$) when $N_{f}$ is about $4$.

Assuming that the transition takes place at a given critical value
of $N_{f}$, we can ask questions about the spectrum of the theory
near the transition. In particular in Ref.~{\cite{AS}}, by studying
Weinberg spectral sum rules, it was observed that for theories
close to a conformal transition parity partners become more
degenerate than in QCD like theories (i.e. theories with $N_{f}$
well below the chiral/conformal transition). This leads to the
possibility that parity doublets can form before chiral symmetry is
actually restored. The lattice studies indicate such a scenario
\cite{mawhinney}.

In Ref.~\cite{ARS}, using a linearly realized effective Lagrangian as a
guide, it was noted that parity doubling is associated with the appearance
of an enhanced global symmetry in the spectrum of the theory. The enhanced
symmetry would develop as the spectrum splits into two sectors, with the
first exhibiting the usual pattern of a spontaneously broken chiral
symmetry, and the second exhibiting an additional, unbroken symmetry and
parity doubling. The first sector includes massless Goldstone bosons and
other states such as massive scalar partners. The second includes a
parity-degenerate vector and axial vector along with other possible parity
partners. The new enhanced symmetry (when the underlying fermions are in a
fundamental and antifundamental representation of the gauge group) has been
identified as either an extra $[SU_{L}(N_{f})\times SU_{R}(N_{f})]$ or a
discrete $Z_{2L}\times Z_{2R}$ one, acting on the massive vector particles.
The underlying $SU_{L}(N_{f})\times SU_{R}(N_{f})$ would affect only the
Goldstones (and their associated chiral partners) and is assumed to break
spontaneously to $SU_{V}(N_{f})$.

The Lagrangian in Ref.~\cite{ARS} thus took the form of a linear $\sigma$
-model coupled to vectors. Since the natural mass scale of this strongly
interacting system is expected to be of order $2\pi v$, where $v$ is the
vacuum expectation value, we could, in principle, include other states as
well. This spectrum choice seemed to be a reasonable approximation when
using the Lagrangian close to a chiral/conformal transition. The massive
degrees of freedom are expected to be very light compared to some intrinsic
scale \cite{Sekhar}. It was noted~\cite{ARS} that if such a near-critical
theory describes symmetry breaking in the electroweak theory, the extra
symmetry suppresses the contribution of the parity doubled sector to the $S$
parameter\cite{PT}. The new dynamical symmetry, hence, plays a key role when
trying to describe a possible strong electroweak Higgs sector \cite{ARS}.

In the linear $\sigma $-model, the scalars appear as the chiral partner of
Goldstone bosons. A different logical possibility is that the lightest
scalar degrees of freedom are not chiral partners of the Goldstone bosons.
This may be the case for QCD (for 3 and 2 light flavors). Indeed here the
lightest scalar nonet could be realized mainly as a four quark \cite{DFSS}
type of meson a l\`{a} Jaffe \cite{Jaffe}. A non linearly realized chiral
Lagrangian might be, in this case, a better phenomenological description of
the low energy physics. Now scalars transform as matter fields under chiral
rotations (see \cite{effective Lagrangians}). While it is certainly an
important issue \cite{DFSS,Jaffe} to understand the true nature of the
scalar bound states as well as the role played by higher order condensates
\cite{ADS,HDC} for strongly interacting theories, here we concentrate on the
symmetries of the effective Lagrangian.

The main aim of this paper is to show the relation between the
enhanced symmetry scenario and a relevant part of any effective
Lagrangian for strongly coupled theories which deals with intrinsic
negative parity terms (i.e. the ones including the Lorentz fully
antisymmetric tensor $\epsilon_{\mu \nu \rho \sigma}$). If such a
near-critical theory is realized in nature the extra symmetry leads
to a distinctive phenomenology associated with these terms.

This relation was not explored in reference \cite{ARS} and anywhere
else in literature. More specifically in Ref.~\cite{ARS} we focused
on the intriguing idea of parity doubling associated with an
enhanced global symmetry and its possible consequences for physics
beyond standard model using a linearly realized Lagrangian.

Another new piece of information in the present article, also not present in
literature, is the explicit construction of the $\epsilon$ term effective
Lagrangian when the underlying fermions are in a pseudoreal representation
of the gauge group.


In this paper we use the non linearly realized effective chiral Lagrangian
to investigate the $\epsilon $ terms (i.e. terms involving the $\epsilon
_{\mu \nu \rho \sigma }$ tensor). We do not expect our results to depend on
this particular choice. To illustrate our results we do not include the
scalars (which can be readily added following Ref.~\cite{effective
Lagrangians}). In this framework the non-abelian global chiral anomalies can
be conveniently encoded in the Wess-Zumino term \cite{Witten,WZ}. The latter
is the first time honored example of {$\epsilon $} term and is needed to
saturate (in the Goldstone phase) at low energies the t'Hooft global anomaly
constraints. At this point the relevant degrees of freedom are the Goldstone
bosons and the massive vector and axial-vector mesons.

We then construct the $\epsilon $ part of the effective Lagrangian
by adding those terms involving spin-1 as well as spin-0 mesons. At
first we formally gauge the Wess-Zumino action, following a
standard procedure developed in \cite{Witten,joe,Schechteretal}.
This procedure automatically provides most of the desired terms
while the local chiral invariance relates the coefficients of the
new $\epsilon$ terms to the Wess-Zumino coefficient. We then
generalize the effective Lagrangian to be only globally invariant
under chiral rotations.

We see that enforcing invariance under the enhanced global symmetry
on the $\epsilon $ part of the Lagrangian, when vectors are
introduced as gauge bosons, requires the absence of the Wess-Zumino
action itself. This is at variance with the t'Hooft global anomaly
constraint (which should be valid for any $N_{f}$ in the broken
phase). However demanding just global invariance for the effective
Lagrangian yields $\epsilon$ terms (describing vector-Goldstone
interactions) no longer related to the Wess-Zumino action, thus
avoiding a topological obstruction to the existence of the enhanced
symmetry.

In particular, we show that if the new enhanced symmetry is a
continuous one (i.e. an extra $SU_{L}(N_{f})\times SU_{R}(N_{f})$
acting only on spin-1 bosons), no dimension four $\epsilon $ term
except the ordinary Wess-Zumino one is allowed. More generally, the
theory splits into two well separated sectors (a sector involves
Goldstone bosons while the other contains the parity doubled
particles). The allowed interactions among the two sectors cannot
be encoded in a single trace and double traces are forbidden at the
4-derivative level. On the other hand if we require invariance only
under a discrete symmetry, which also guarantees parity doubling, a
few dimension four $\epsilon $ terms (specifically 3 for the
$SU_{L}(N_{f})\times SU_{R}(N_{f})$ case) survive. The enhanced
symmetry leads to a distinctive phenomenology associated with the
intrinsic negative parity terms.

It is worth noting that the effective Lagrangian employed here, while
describing the global symmetries, does not accurately describe the dynamics
of a chiral/conformal transition. That is, it cannot be used directly as the
basis for a Landau-Ginzburg theory of this transition with its expected
nonanalytic behavior \cite{ATW}. In Ref.~\cite{SS}, an approach to such a
Landau-Ginzburg theory was developed to describe the usual global
symmetries. This approach could perhaps be extended to include the vectors
of the present effective Lagrangian. We expect that it would describe the
same symmetries we have considered here, both the spontaneously broken
symmetry and the additional, unbroken symmetry.

In Section \ref{nlrs} we construct the non $\epsilon $ term
effective Lagrangian using a non linearly realized underlying
global symmetry $SU_{L}(N_{f})\times SU_{R}(N_{f})$. Parity
invariance is imposed along with the ordinary pattern of chiral
symmetry breaking (i.e. $SU_{L}(N_{f})\times
SU_{R}(N_{f})\rightarrow SU_{V}(N_{f})$). We have $N_{f}^{2}-1$
Nambu-Goldstone bosons, and complete the low energy spectrum with a
set of vector and axial vector fields. Then, briefly reviewing the
analysis done in Ref.~\cite{ARS}, we investigate the spectrum and
show that there is a particular choice of the parameters in the
theory which allows for a degenerate vector and axial vector
spectrum while augmenting the global symmetry to include an extra
$SU_{L}(N_{f})\times SU_{R}(N_{f})$. The possible appearance of an
additional continuous symmetry was considered by Casalbuoni {\it et
al.} in Ref.\cite{sei}. These papers were restricted to the case
$N_{f}=2$ and did not include discussion of the possible connection
to a near critical theory. As already observed in Ref.~ \cite{ARS}
we note that a smaller discrete symmetry $Z_{2L}\times Z_{2R}$ is
also capable to guarantee parity doubling. In Section III we
introduce the Wess-Zumino action. By gauging it we generate most of
the $\epsilon $ terms consistent with local chiral invariance,
parity and charge conjugation. We then generalize the $\epsilon $
Lagrangian to be globally invariant, and investigate its symmetry
properties. We also discuss some phenomenological aspects related
to the intrinsic negative parity terms. In Section IV and V we
extend the study to fundamental fermions in a pseudoreal
representation of the underlying gauge group. The pseudoreal
representations allow for the lowest number of colors (i.e. $N=2$)
and consequently for the lowest possible number of flavors for
which the theory might show a dynamically enhanced symmetry. The
enhanced global symmetry is $\left[ SU(2N_{f})\right]^{2}$
spontaneously broken to $Sp(2N_{f})\times SU(2N_{f})$. In Section
VI we conclude. Appendix A contains some detailed proofs and
computations not present in the main text.

\section{Effective Lagrangian for $SU(N_f)\times SU(N_f)$ global symmetry}

\label{nlrs}

We construct an effective Lagrangian which manifestly possesses the
global symmetry $SU_L(N_f)\times SU_R(N_f)$ of the underlying
theory. We assume that chiral symmetry is broken according to the
standard pattern $SU_L(N_f)\times SU_R(N_f)\rightarrow SU_V(N_f)$.
The $N_f^2 - 1$ Goldstone bosons are encoded in the $N_f\times N_f$
matrix $U$ transforming linearly under a chiral rotation (the
matrix $U$ replaces $M$ in Ref. \cite{ARS} for non linear
realizations):
\begin{equation}
U\rightarrow u_L U u_R^{\dagger} \ ,
\end{equation}
with $u_{L/R} \in SU_{L/R}(N_f)$. $U$ satisfies the non linear realization
constraint $U U^{\dagger} =1 $. We also require ${\rm det} U =1$. In this
way we avoid discussing the axial $U_A(1)$ anomaly at the effective
Lagrangian level (see Ref.~{\cite{SS}} for a general discussion of
anomalies). We have
\begin{equation}
{U=e^{i \frac{\Phi}{v}}} \ ,
\end{equation}
with $\Phi=\sqrt{2}\Phi^a T^a$ representing the $N^2_f - 1$
Goldstone bosons. $T^{a}$ are the generators of $SU(N_{f})$, with
$a=1,...,N_{f}^{2}-1$ and $\displaystyle{{\rm Tr}\left[
T^{a}T^{b}\right] = \frac{1}{2}\delta ^{ab}}$. $v$ is the vacuum
expectation value.

We enlarge the spectrum of massive particles including vector and
axial-vector fields as in Ref.\cite{ARS,joe}. This is formally done by
imposing parity invariance and gauging the chiral group through the
introduction of the covariant derivative
\begin{equation}
D^{\mu }U=\partial ^{\mu }U-iA_{L}^{\mu }U+iUA_{R}^{\mu }\ ,
\end{equation}
where $A_{L/R}^{\mu }=A_{L/R}^{\mu ,a}T^{a}$\footnote{%
We rescale $A$ by the coupling constant $\tilde{g}$ with respect to the
notation in Ref \cite{ARS}.}. Under a chiral transformation
\begin{equation}
A_{L/R}^{\mu }=u_{L/R}A_{L/R}^{\mu }u_{L/R}^{\dagger }-i\partial ^{\mu
}u_{L/R}u_{L/R}^{\dagger }.  \label{gaugetrans}
\end{equation}

The effective Lagrangian including vectors and globally invariant under
chiral rotations is (up to two derivatives and counting $U$ as a
dimensionless field)
\begin{eqnarray}
L=~ &&\frac{v^{2}}{2}\,{\rm Tr}\left[ D_{\mu }UD^{\mu }U^{\dagger }\right]
+m^{2}\,{\rm Tr}\left[A_{L\mu }A_{L}^{\mu }+A_{R\mu }A_{R}^{\mu }\right]
\nonumber \\
&+&hv^{2}\,{\rm Tr}\left[ A_{L\mu }UA_{R}^{\mu }U^{\dagger }\right] +i\,{s}
v^{2}\,{\rm Tr}\left[ A_{L\mu }\left( UD^{\mu }U^{\dagger }\right) +A_{R\mu
}\left( U^{\dagger }D^{\mu }U\right) \right] \ .  \label{nr1}
\end{eqnarray}
We also count $A_{L/R}$ as derivative and impose parity. The parameters $h$
and $s$ are dimensionless and real, while $m$ is the common spin-$1$ mass.

One big advantage of using non linear realizations is that the
bosonic potential is absent. A similar Lagrangian has been derived
in Ref.~{\cite {sei}} using the hidden gauge symmetry method
{\cite{BKY}}. Another well known difference with respect to the
linear realizations is the absence of the scalars as chiral
partners of the Goldstone bosons. This is, in general, physically
acceptable only if the scalars are heavier than all the relevant
degrees of freedom considered in the Lagrangian. In the event they
have a mass comparable to the massive degrees of freedom, already
considered, we can include them following Ref.~\cite{effective
Lagrangians}. We finally add to the Lagrangian a standard kinetic
term for the vector fields,
\begin{equation}
L_{Kin}=-\frac{1}{2\tilde{g}^{2}}{\rm Tr}\left[ F_{L\mu \nu }F_{L}^{\mu \nu
}+F_{R\mu \nu }F_{R}^{\mu \nu }\right] \ ,  \label{KF}
\end{equation}
where
\begin{equation}
F_{L/R}^{\mu \nu }=\partial ^{\mu }A_{L/R}^{\nu }-\partial ^{\nu
}A_{L/R}^{\mu }-i\left[ A_{L/R}^{\mu },A_{L/R}^{\nu }\right] \ .  \label{F}
\end{equation}
This effective Lagrangian, with its massive vectors and axial
vectors, is of course not renormalizable but it can nevertheless
provide a reasonable description of low-lying states\footnote{A
Lagrangian of this type plays this role for the low-lying QCD
resonances~ \cite{effective Lagrangians}.}. While it cannot be a
complete description of the hadronic spectrum, it has sufficient
content to guide a general discussion of enhanced symmetries.

Keeping only terms quadratic in the fields, the Lagrangian in Eq.
(\ref{nr1}) takes the form
\begin{equation}
L=~\frac{1}{2}{\rm Tr}\left[ \partial _{\mu }\Phi \partial ^{\mu
}\Phi
\right] +\sqrt{2}(s-1)\,{v}\,{\rm Tr}\left[ \partial _{\mu }\Phi A^{\mu }
\right] +M_{A}^{2}{\rm Tr}\left[ A_{\mu }A^{\mu }\right] +M_{V}^{2}{\rm Tr}
\left[ V_{\mu }V^{\mu }\right] \ ,  \label{quadratic}
\end{equation}
where we have defined the new vector fields
\begin{equation}
V=\frac{A_{L}+A_{R}}{\sqrt{2}}\ ,\qquad \qquad A=\frac{A_{L}-A_{R}}{\sqrt{2}}
\ .  \label{va}
\end{equation}
The vector and axial masses are related to the effective Lagrangian
parameters via
\begin{eqnarray}
M_{A}^{2} &=&m^{2}+v^{2}\,\left[ 1-2\,s\,-\frac{h}{2}\right] \,,  \nonumber
\\
M_{V}^{2} &=&m^{2}+v^{2}\frac{h}{2}\,.
\end{eqnarray}

The second term in Eq.~(\ref{quadratic}) mixes the axial vector with the
Goldstone bosons. This kinetic mixing may be diagonalized away by the field
redefinition
\begin{equation}
A\rightarrow A+v\frac{1-s}{\sqrt{2}M_{A}^{2}}\partial \Phi \ ,
\end{equation}
leaving the mass spectrum unchanged~\cite{joe}. The axial vector mass
difference is given by
\begin{equation}
M_{A}^{2}-M_{V}^{2}=v^{2}\,\left[ 1-2s-h\right] \ .  \label{differenza}
\end{equation}
In QCD this difference is known experimentally to be positive, a fact that
can be understood by examining the Weinberg spectral function sum rules (see
Ref.~{\cite{BDLW}} and references therein). The effective Lagrangian
description is of course less restrictive. Depending on the values of the $s$
and $h$ parameters, one can have a degenerate or even inverted mass
spectrum. In this case, it is clear that an underlying theory which could
provide such a scenario has to be different from QCD, allowing for the
modification of the spectral function sum rules.

In Ref. \cite{ARS} it was suggested that an enhanced symmetry, which would
provide a parity-doubled spectrum, could arise as the low energy limit of an
underlying quasi-conformal theory. This enhanced symmetry requires a
particular relation among the parameters of the theory. In non linear
realizations the relation reads:
\begin{equation}
s=1,\qquad h=-1\ .
\end{equation}
The quadratic Lagrangian in Eq.(\ref{quadratic}) can then be written as:
\begin{equation}
L=~\frac{1}{2}{\rm Tr}\left[ \partial _{\mu }\Phi \partial ^{\mu
}\Phi
\right] +M^{2}{\rm Tr}\left[ A_{\mu L}A_{L}^{\mu }+A_{\mu R}A_{R}^{\mu }
\right] \,  \label{22}
\end{equation}
where $M^{2}=m^{2}-\frac{v^{2}}{2}$ and the global symmetry is
enhanced to $\left[ SU_{L}(N_{f})\times SU_{R}(N_{f})\right] ^{2}$
spontaneously broken by the vacuum to $SU_{V}(N_{f})\times \left[
SU_{L}(N_{f})\times SU_{R}(N_{f})\right] $.

Besides the extra global symmetry $SU_L(N_f)\times SU_R(N_f)$ the
Lagrangian in Eq.~(\ref{22}) possesses an extra discrete
$Z_{2L}\times Z_{2R}$ symmetry (which, alone, is enough to protect
the axial vector degeneracy). Under $Z_{2L}\times Z_{2R}$ the
vector fields transform according to
\begin{equation}
A_L \rightarrow z_{L} A_L \ , \qquad A_R \rightarrow z_{R}A_R \ ,
\end{equation}
with $z_{L/R}=1,-1$ and $z_{L/R}\in Z_{2L/R}$. Hence, at the two derivative
level (for non linear realizations), the extra symmetry is $Z_{2L}\times
Z_{2R}\times \left[ SU_{L}(N_{f})\times SU_{R}(N_{f})\right] $.

A comment is in order here. Now the standard kinetic term for the
vectors in Eq.~(\ref{KF}) entering at the four derivative level
breaks the discrete symmetry through the presence of the triple
vector interaction. If this symmetry is the one chosen by nature we
can use a new $F_{L/R}^{\mu \nu }=\partial^{\mu }A_{L/R}^{\nu
}-\partial ^{\nu }A_{L/R}^{\mu }$ to replace the one in
Eq.~(\ref{KF}) while we still have the freedom to add quartic
interactions in the vector field.

There is no interaction term between spin one fields and the
Goldstones surviving at the two derivative level. However such
terms do emerge at the four derivative level. If we insist on a new
continuous enhanced global symmetry governing the parity doubling
physics it is easy to see that the most general interaction terms
allowed are:
\begin{eqnarray}
L_{4}=&+&a_1 {\rm Tr}\left[\partial_{\mu}U
\partial^{\mu}U^{\dagger}\right]\, {\rm Tr} \left[A_{L\nu}
A_L^{\nu} + A_{R\nu} A_R^{\nu} \right]  \nonumber \\ &+&a_2 {\rm
Tr}\left[\partial_{\mu}U \partial_{\nu}U^{\dagger}\right]\,{\rm Tr}
\left[A_L^{\mu} A_L^{\nu} + A_R^{\mu} A_R^{\nu} \right] \ ,
\label{four}
\end{eqnarray}
with $a_1$ and $a_2$ real. The continuous enhanced symmetry forces
the interactions among the Goldstones and vectors to appear via
double traces. To complete the previous 4-derivative Lagrangian we
should add the, well known, 4-derivative terms describing the
Golstone self-interactions at this order \cite{GL} and include, as
well, the vector self-interactions involving up to four spin one
fields.

It is interesting to note that if parity doubling is associated
just to an invariance under the discrete $Z_{2L}\times Z_{2R}$
symmetry we also have single trace interaction terms, like for
example:
\begin{equation}
{\rm Tr}\left[\partial_{\mu}U \partial^{\mu}U^{\dagger}
\left(A_{L\nu} A_L^{\nu} + A_{R\nu} A_R^{\nu} \right) \right] \ ,
\quad {\rm Tr}\left[A_{L\mu} A_L^{\mu} U A_{R\nu} A_R^{\nu}
U^{\dagger} \right] \ .
\end{equation}
We can then disentangle, at the four derivative level, which
enhanced global symmetry protects the vector-axial mass degeneracy.

The generalization to the electroweak sector can be easily followed by using
the procedure outlined in Ref \cite{ARS}. We have seen in Ref \cite{ARS}
that the enlarged symmetry works as partial custodial symmetry for the $S$
parameter (defined in \cite{PT}) when describing electroweak symmetry
breaking via strong interactions. This intriguing feature has been first
explored in Ref.~\cite{sei}.

\section{The ${\protect\epsilon}$ terms for $SU(N_f)\times SU(N_f)$}

In the previous section, we have shown that interaction terms among
Goldstone bosons and vectors appear at the 4-derivative level when
discussing a theory with a parity doubled spectrum, and we have
limited our discussion to intrinsic parity even interaction terms.
Another important set of 4-derivative terms are the intrinsic
parity odd terms involving spin one and spin zero fields. These
terms contain the Lorentz antisymmetric tensor $\epsilon_{\mu \nu
\rho \sigma}$.
The Wess-Zumino action is the first example of $\epsilon$ term. It can be
compactly written using the language of differential forms. It is useful to
introduce the Maurer-Cartan one forms:
\begin{equation}
\alpha=\left(\partial_{\mu}U\right)U^{-1}\, dx^{\mu}\equiv
\left(dU\right)U^{-1}\ , \quad \beta=U^{-1}dU=U^{-1}\alpha U \ .  \label{MC}
\end{equation}
$\alpha$ and $\beta$ are algebra valued one forms and transform,
respectively, under the left and right $SU(N_f)$ flavor group. The
Wess-Zumino effective action is
\begin{equation}
\Gamma_{WZ}\left[U\right]=C\, \int_{M^5} {\rm Tr} \left[\alpha^5\right] \ .
\label{WZ}
\end{equation}
The price to pay in order to make the action local is to augment by one the
space dimensions. Hence the integral must be performed over a
five-dimensional manifold whose boundary ($M^4$) is the ordinary Minkowski
space. The constant $C$ is fixed to be
\begin{equation}
C=-i\frac{N}{240\pi^2} \ ,
\end{equation}
by comparing the current algebra prediction for the time honored
process $\pi^{0}\rightarrow 2\gamma$ with the amplitude predicted
using Eq.~(\ref{WZ}) once we gauge the electromagnetic sector of
the Wess-Zumino term, and $N$ is the number of colors.

We now consider $\epsilon$ type terms involving the vector and
axial vector particles. As for the non $\epsilon$ part of the
Lagrangian, discussed in Section~\ref{nlrs}, we first gauge the WZ
term under the $SU_L(N_f)\times SU_R(N_f)$ chiral symmetry group.
This procedure automatically induces new $\epsilon$ terms
\cite{Witten,joe,Schechteretal}, leading to the following
Lagrangian,
\begin{eqnarray}
\Gamma _{WZ}\left[ U,\;A_{L},\;A_{R}\right] &=&\Gamma _{WZ}\left[ U\right]
\,+\,5Ci\,\int_{M^{4}}{\rm Tr}\left[ A_{L}\alpha ^{3}+A_{R} \beta ^{3}\right]
\nonumber \\
&&-5C\,\int_{M^{4}}{\rm Tr}\left[ (dA_{L}A_{L}+A_{L}dA_{L})\alpha
+(dA_{R}A_{R}+A_{R}dA_{R})\beta \right]  \nonumber \\
&&+5C\,\int_{M^{4}}{\rm Tr}\left[
dA_{L}dUA_{R}U^{-1}-dA_{R}dU^{-1}A_{L}U \right]  \nonumber \\
&&+5C\,\int_{M^{4}}{\rm Tr}\left[ A_{R}U^{-1}A_{L}U\beta
^{2}-A_{L}UA_{R}U^{-1}\alpha ^{2}\right]  \nonumber \\
&&+\frac{5C}{2}\,\int_{M^{4}}{\rm Tr}\left[ (A_{L}\alpha
)^{2}-(A_{R}\beta )^{2}\right] +5Ci\,\int_{M^{4}}{\rm Tr}\left[
A_{L}^{3}\alpha +A_{R}^{3}\beta \right]  \nonumber \\
&&+5Ci\,\int_{M^{4}}{\rm Tr}\left[
(dA_{R}A_{R}+A_{R}dA_{R})U^{-1}A_{L}U-(dA_{L}A_{L}
+A_{L}dA_{L})UA_{R}U^{-1}
\right]  \nonumber \\
&&+5Ci\,\int_{M^{4}}{\rm Tr}\left[ A_{L}UA_{R}U^{-1}A_{L}\alpha
+A_{R}U^{-1}A_{L}UA_{R}\beta \right]  \nonumber \\
&&+5C\,\int_{M^{4}}{\rm Tr}\left[
A_{R}^{3}U^{-1}A_{L}U-A_{L}^{3}UA_{R}U^{-1}+\frac{1}{2}
(UA_{R}U^{-1}A_{L})^{2}\right]  \nonumber \\
&&-5Cr\,\int_{M^{4}}{\rm Tr}\left[ F_{L}UF_{R}U^{-1}\right] \ .  \label{GWZ1}
\end{eqnarray}
Here the two-forms $F_{L}$ and $F_{R}$ are defined as
$F_{L}=dA_{L}-iA_{L}^{2}$ and $F_{R}=dA_{R}-iA_{R}^{2}$ with the
one form $A_{L/R}=A^{\mu}_{L/R}dx_{\mu}$. The previous Lagrangian,
when identifying the vector fields with true gauge vectors,
correctly saturates the underlying global anomalies.

The last term in Eq.~(\ref{GWZ1}) is a gauge covariant term which can always
be added if parity is not imposed. {}For reader's convenience we provide the
rules for parity transformation
\begin{eqnarray}
&&A_{L,R}(\vec{x})\leftrightarrow A_{R,L}(-\vec{x})\;,\qquad
U(\vec{x} )\leftrightarrow U^{-1}(-\vec{x})\;,  \nonumber \\
&&\alpha \leftrightarrow -\beta \;,\qquad d\leftrightarrow
d\;,\qquad \mbox{(measure)}\leftrightarrow -\mbox{(measure)}\ ,
\label{parity}
\end{eqnarray}
as well as for charge conjugation
\begin{equation}
A_{L,R}\leftrightarrow -A_{R,L}^{T}\;,\quad U\leftrightarrow U^{T}\;,\quad
\alpha \leftrightarrow \beta ^{T}\;.  \label{charge}
\end{equation}
The last term in Eq.(\ref{GWZ1}) is not invariant under parity, so the
parameter $r$ must vanish. All the other terms are related by gauge
invariance.

Imposing just global chiral invariance, together with $P$ and $C$, the
previous Lagrangian has ten unrelated terms:
\begin{eqnarray}
\Gamma _{WZ}\left[ U,\;A_{L},\;A_{R}\right] &=&\Gamma _{WZ}\left[ U\right]
\,+\,5c_{1}\,i\int_{M^{4}}{\rm Tr}\left[ A_{L}\alpha ^{3}+A_{R}\beta ^{3}
\right]  \nonumber \\
&&+5c_{2}\,\int_{M^{4}}{\rm Tr}\left[
(dA_{L}A_{L}+A_{L}dA_{L})\alpha +(dA_{R}A_{R}+A_{R}dA_{R})\beta
\right]  \nonumber \\ &&-5c_{3}\int_{M^{4}}{\rm Tr}\left[
dA_{L}dUA_{R}U^{-1}-dA_{R}dU^{-1}A_{L}U
\right]  \nonumber \\
&&-5c_{4}\,\int_{M^{4}}{\rm Tr}\left[ A_{R}U^{-1}A_{L}U\beta
^{2}-A_{L}UA_{R}U^{-1}\alpha ^{2}\right]  \nonumber \\
&&-\frac{5c_{5}}{2}\int_{M^{4}}{\rm Tr}\left[ (A_{L}\alpha
)^{2}-(A_{R}\beta )^{2}\right] +5c_{6}\,i\int_{M^{4}}{\rm Tr}\left[
A_{L}^{3}\alpha +A_{R}^{3}\beta \right]  \nonumber \\
&&+5c_{7}\,i\int_{M^{4}}{\rm Tr}\left[
(dA_{R}A_{R}+A_{R}dA_{R})U^{-1}A_{L}U-(dA_{L}A_{L}+A_{L}dA_{L})UA_{R}U^{-1}
\right]  \nonumber \\
&&+5c_{8}\,i\int_{M^{4}}{\rm Tr}\left[ A_{L}UA_{R}U^{-1}A_{L}\alpha
+A_{R}U^{-1}A_{L}UA_{R}\beta \right]  \nonumber \\
&&-5c_{9}\,\int_{M^{4}}{\rm Tr}\left[
A_{R}^{3}U^{-1}A_{L}U-A_{L}^{3}UA_{R}U^{-1}\right]  \nonumber \\
&&- \frac{5c_{10}}{2} \,\int_{M^{4}}{\rm Tr}\left[
(UA_{R}U^{-1}A_{L})^{2}
\right] \ ,  \label{GWZ2}
\end{eqnarray}
\noindent where the $c$-coefficients are imaginary. It is for this rather
large arbitrariness as well as vector meson dominance considerations that,
at times, in literature (for QCD) \cite{joe}, one reduces the parameter
space by using the locally invariant Lagrangian rather than the globally
invariant one.

While the gauging procedure of the Wess Zumino term automatically generates
a large number of $\epsilon $ terms, it does not guarantee that we have
uncovered all terms consistent with chiral, $P$ and $C$ invariance. Indeed
there is still one new single trace term, not present in literature, to add
to the action:
\begin{equation}
c_{11}i\int_{M^{4}}{\rm Tr}\left[ A_{L}^{2}\left( UA_{R}U^{-1}\alpha -\alpha
UA_{R}U^{-1}\right) +A_{R}^{2}\left( U^{-1}A_{L}U\beta -\beta
U^{-1}A_{L}U\right) \right] \ ,  \label{undici}
\end{equation}
and $c_{11}$ is an imaginary coefficient. Imposing invariance under $CP$ has
been very useful to reduce the number of possible $\epsilon $ terms. For
example it is easy to verify that a term of the type ${\rm Tr}\left[
dA_{L}\left( UA_{R}U^{-1}\right) ^{2}\right] $ is $CP$ odd.

In Appendix A we provide a general proof that all the dimension four (i.e.
4-derivative) terms involving the Lorentz tensor $\epsilon_{\mu \nu \rho
\sigma}$, which are consistent with global chiral symmetries as well as $C$
and $P$ invariance, are the ones presented in Eq.~(\ref{GWZ2}) and
Eq.~(\ref {undici}).

We see that the Wess-Zumino term, containing only Goldstones, does not
interfere with new possible symmetries involving the spin-1 spectrum.

If the theory develops the extra $SU_L(N_f)\times SU_R(N_f)$ when acting
over the vector states, the continuous enhanced global symmetry will allow
only the standard Wess-Zumino term and all the $c$-coefficients {\it must
vanish identically}.

Hence we discover that while even parity four derivative
Goldstone-vector interaction terms are allowed by the continuous
enhanced symmetry (see Eq.~(\ref{four})), parity odd, four
derivative terms are forbidden.

This fact implies that $\epsilon$ type interactions between vectors and
Goldstone bosons must be higher order in the derivative expansion. For
example we can have a term like:
\begin{equation}
\Gamma_{WZ} \times {\rm Tr} \left[A_{L\mu}A^{\mu}_{L}+ A_{R\mu}A^{\mu}_{R}\right] \ .
\end{equation}
\noindent In general the continuous enhanced global symmetry does not allow
mixing between the Goldstone bosons and the spin-1 sector via a single trace.

In the case we associate the parity doubled spectrum only to a
$Z_{2L}\times Z_{2R}$ discrete symmetry, besides the Wess-Zumino
term, we still have non vanishing terms, namely, $c_{2}$, $c_{5}$
and $c_{10}$.

We have seen that the enhanced global symmetry imposes strong
constraints on even, as well as, odd intrinsic parity terms of a
general effective Lagrangian describing the strong dynamics of any
underlying gauge theory displaying a parity doubled spectrum. In
particular the phenomenology of such a theory is predicted to be
quite different from the one observed in QCD with 2 or 3 flavors.

The $\epsilon$ terms \cite{joe,Schechteretal} play an important
role in QCD and are essential to correctly describe the vector
meson decays and as a stabilizing piece (which can replace the
Skyrme term) in the solitonic picture of the nucleon. A striking
phenomenological difference with respect to QCD, due to the low
energy enhanced global symmetry, is the {\it absence} of vector
meson decay into three Goldstone bosons. When no explicit chiral
symmetry breaking terms are present the vectors are stable.
Interestingly, when the new global symmetry is $Z_{2L}\times
Z_{2R}$, the $c_2$ term survives leading to a single pion plus two
vector interaction. We stress that these processes are
characteristic of the $\epsilon$ structure of the interactions and
can be used as clear phenomenological signals of a parity doubling
physics.

If such a near-critical theory describes symmetry breaking in the
electroweak theory, the extra symmetry, not only suppresses the
contribution of the parity doubled sector to the $S$ parameter as
shown in \cite{ARS}, but also leads to a distinctive phenomenology
associated with the intrinsic negative parity terms.

\section{Effective Lagrangian for $SU(2N_f)$ global symmetry}

In this section we consider the interesting case of fermions in pseudoreal
representations of the gauge group. The simplest example is offered by an
underlying $SU(2)$ gauge theory, a choice that will also provide the
smallest value for the critical $N_f$~\cite{postmodern}. However our
formalism does not restrict to the case $N=2$ but can be applied to any
underlying theory with fermions in a pseudoreal representation of the gauge
group with given (larger number of flavors) global flavor symmetry. Such
theories are currently being investigated on the lattice (see Ref.~\cite
{Mariapaola}). The quantum global symmetry for $N_f$ matter fields in the
pseudoreal representation of the gauge group~\cite{PT} is $SU(2N_f)$ which
contains $SU_L(N_f)\times SU_R(N_f)$. We expect the gauge dynamics to create
a non vanishing fermion-antifermion condensate which breaks the global
symmetry to $Sp(2 N_f)$.

We divide the generators ${T}$ of $SU(2N_f)$, normalized according
to $\displaystyle{{\rm
Tr}\left[T^aT^b\right]}=\frac{1}{2}\delta^{ab}$, into two classes.
We call the generators of $Sp(2N_f)$ $\{S^{a}\}$ with $a=1,\ldots
,2N_f^2+N_f$, and the remaining $SU(2N_{f})$ generators
(parameterizing the quotient space $SU(2N_{f})/Sp(2N_{f})$)
$\{X^{i}\}$ with $i=1,\ldots ,2N_f^2-N_f-1$.

This breaking pattern gives $2N^2_f - N_f -1$ Goldstone bosons, encoded in
the antisymmetric matrix $U^{ij}$ and $i,j = 1,\ldots,2N_f$ as follows:
\begin{equation}
U=e^{i\frac{\Pi ^{i}X^{i}}{v}}\,E\ ,
\end{equation}
where the $N_f\times N_f$ matrix $E$ is
\begin{equation}
E=\left(
\begin{array}{cc}
{\bf 0} & {\bf 1} \\
-{\bf 1} & {\bf 0}
\end{array}
\right) \ .
\end{equation}
$U$ transforms linearly under a chiral rotation
\begin{equation}
U\rightarrow u\, U \, u^{T} \ ,
\end{equation}
with $u\in SU(2N_f)$. The non linear realization constraint,
$\displaystyle {UU^{\dagger }=1}$, is automatically satisfied.

The generators of the $Sp(2N_f)$ satisfy the following relation,
\begin{equation}
S^{T}\,E+E\,S=0\ ,
\end{equation}
while the $X^i$ generators obey,
\begin{equation}
X^{T}=E\,X\,E^{T}\ ,
\end{equation}
Using this last relation we can easily demonstrate that $U^{T}=-U$. We also
require
\begin{equation}
{\rm Pf}\,U=1\ ,
\end{equation}
avoiding in this way to consider the explicit realization of the underlying
axial anomaly at the effective Lagrangian level.

We define the following vector field
\begin{equation}
A_{\mu }=A_{\mu }^{a}T^{a}\ ,
\end{equation}
which formally transforms under a $SU(2N_{f})$ rotation as
\begin{equation}
A_{\mu }\rightarrow uA_{\mu }u^{\dagger }-i\partial _{\mu }uu^{\dagger }\ .
\end{equation}

Following the procedure outlined in the previous sections, we can define a
formal covariant derivative as
\begin{equation}
D_{\mu }U=\partial _{\mu }U-iA_{\mu }U-iUA_{\mu }^{T}\ .
\end{equation}
The non linearly realized effective Lagrangian up to two derivatives and
invariant under an $SU(2N_{f})$ transformation is
\begin{eqnarray}
L=~ &&v^{2}\,{\rm Tr}\left[ D_{\mu }UD^{\mu }U^{\dagger }\right] +m^{2}\,
{\rm Tr}\left[ A_{\mu }A^{\mu }\right]  \nonumber \\
&+&hv^{2}\,{\rm Tr}\left[ A_{\mu }UA^{T\mu }U^{\dagger }\right] +i\,sv^{2}\,
{\rm Tr}\left[ A_{\mu }UD^{\mu }U^{\dagger }\right] \ .  \label{nr2}
\end{eqnarray}
By investigating the spectrum, as for $SU_L(N_{f})\times
SU_R(N_{f})$, we see \cite{ARS} that the global symmetry group
becomes $Sp(2N_{f})\times \left[SU(2N_{f})\right] $ for
\begin{equation}
s=4\ ,\qquad h=2\ .
\end{equation}
The extra $SU(2N_{f})$ acts only on the vector field, and the associated
effective Lagrangian is:
\begin{equation}
L=v^{2}\,{\rm Tr}\left[ \partial _{\mu }U\partial ^{\mu }U^{\dagger }\right]
+M^{2}{\rm Tr}\left[ A_{\mu }A^{\mu }\right] \ ,
\end{equation}
where $M^{2}=m^{2} - 2 v^{2}$. The Lagrangian also possesses the extra
global $Z_{2}$ (i.e. $A\rightarrow zA$, with $z=\pm 1$) symmetry.

We complete the effective Lagrangian by adding the kinetic term for the
vectors:
\begin{equation}
L_{Kin}=-\frac{1}{2\tilde{g}^{2}}{\rm Tr}\left[ F_{\mu \nu }F^{\mu
\nu } \right] \ ,
\end{equation}
where
\begin{equation}
F^{\mu \nu }=\partial ^{\mu }A^{\nu }-\partial ^{\nu }A^{\mu
}-i\left[ A^{\mu },A^{\nu }\right] \ ,
\end{equation}
which for the $Z_2$ enhanced symmetry case we simply write as
$F^{\mu
\nu }=\partial ^{\mu }A^{\nu }-\partial ^{\nu }A^{\mu }$.

 As for the $SU_L(N_f)\times SU_R(N_F)$ case, interactions
between vectors and Goldstones arise naturally through double trace
terms. For the continuous $SU(2N_f)$ extra symmetry case:
\begin{equation}
L_{4}=+a_1 {\rm Tr}\left[\partial_{\mu}U
\partial^{\mu}U^{\dagger}\right]\, {\rm Tr} \left[A_{\nu}
A^{\nu}\right] + a_2 {\rm Tr}\left[\partial_{\mu}U
\partial_{\nu}U^{\dagger}\right]\,{\rm Tr} \left[A^{\mu} A^{\nu}\right] \ ,
\label{foursp}
\end{equation}
with $a_1$ and $a_2$ real coefficients and for the $Z_{2L}\times
Z_{2R}$ symmetry through, for example, single trace terms of the
type:
\begin{equation}
{\rm Tr}\left[\partial_{\mu}U \partial^{\mu}U^{\dagger}A_{\nu}
A^{\nu}\right] \ , \quad {\rm Tr}\left[A_{\mu} A^{\mu} U A_{\nu}^T
A^{T\nu} U^{\dagger} \right] \ .
\end{equation}

\section{The ${\protect\epsilon}$ terms for $SU(2N_{f})$}

In this section we generate the $\epsilon $ terms following the same
procedure used for the $SU_{L}(N_{f})\times SU_{R}(N_{f})$ global symmetry
case. First we introduce the one form
\begin{equation}
\alpha =\left( dU\right) U^{-1}\ .
\end{equation}
It is sufficient to define only $\alpha $ since the analog of $\beta
=U^{-1}dU=\alpha ^{T}$ is now not an independent form. The Wess-Zumino
action term is:
\begin{equation}
\tilde{\Gamma}_{WZ}\left[ U\right] =C\,\int_{M^{5}}{\rm Tr}\left[ \alpha
^{5} \right] \ ,  \label{WZSp}
\end{equation}
where again we are integrating on a five dimensional manifold and $C=-i\frac{
2}{240\pi ^{2}}$ for $N=2$.

We now gauge the Wess-Zumino action under the $SU(2N_f)$ chiral symmetry
group. This procedure provides single trace $\epsilon$-terms involving
vector, axial and Goldstones with an universal coupling $C$. The gauged $CP$
invariant Wess-Zumino term is
\begin{eqnarray}
\tilde{\Gamma}_{WZ}\left[ U,\;A\right] &=&\tilde{\Gamma}_{WZ}\left[ U\right]
\,+\,10Ci\int_{M^{4}}{\rm Tr}\left[ A\alpha ^{3}\right]  \nonumber \\
&&-10C\,\int_{M^{4}}{\rm Tr}\left[ (dAA+AdA)\alpha \right]  \nonumber \\
&&-5C\,\int_{M^{4}}{\rm Tr}\left[ dAdUA^{T}U^{-1}-dA^{T}dU^{-1}AU \right]
\nonumber \\
&&-5C\,\int_{M^{4}}{\rm Tr}[UA^{T}U^{-1}(A\alpha ^{2}+\alpha ^{2}A)]
\nonumber \\
&&+\,5C\int_{M^{4}}{\rm Tr}\left[ (A\alpha )^{2}\right] +10C\,i\int_{M^{4}}
{\rm Tr}\left[ A^{3}\alpha \right]  \nonumber \\
&&+10C\,i\int_{M^{4}}{\rm Tr}\left[ (dAA+AdA)UA^{T}U^{-1}\right]  \nonumber
\\
&&-10C\,i\int_{M^{4}}{\rm Tr}\left[ A\alpha AUA^{T}U^{-1}\right]  \nonumber
\\
&&+10C\,\int_{M^{4}}{\rm Tr}\left[ A^{3}UA^{T}U^{-1}+\frac{1}{4}
(AUA^{T}U^{-1})^{2}\right] \ ,
\end{eqnarray}
where $A=A^{\mu}dx_{\mu}$. We soon realize that unless $C$ vanishes
identically this Lagrangian does not possesses an enhanced symmetry. However
a vanishing $C$ is forbidden by the global t'Hooft anomaly matching
constraint. We need to disentangle the Wess-Zumino term from the other $%
\epsilon$ interactions. This is readily realized by generalizing the
Lagrangian to be globally invariant under a chiral rotation and invariant
under $CP$.

It is useful to define the following parity and charge conjugate
transformations. For parity we have:
\begin{eqnarray}
&&
\begin{array}{cc}
A(\vec{x})\rightarrow -E^{T}A^{T}(-\vec{x})E\;,\quad & U(\vec{x}
)\leftrightarrow -U^{-1}(-\vec{x})\;,
\end{array}
\nonumber \\
&&
\begin{array}{ccc}
\alpha \leftrightarrow -\alpha ^{T}\;,\quad & d\leftrightarrow d\;,\quad &
\mbox{(measure)}\leftrightarrow -\mbox{(measure)}\ .
\end{array}
\end{eqnarray}
For charge conjugation:
\begin{equation}
A\rightarrow {\cal C}A{\cal C}^{\dagger}\ , \quad U\rightarrow
-{\cal C}U {\cal C}^{\dagger} \ , \quad \alpha\rightarrow {\cal
C}\alpha {\cal C}
^{\dagger} \ ,
\end{equation}
with
\begin{equation}
{\cal C}=\left(
\begin{array}{cc}
{\bf 0} & {\bf 1} \\
{\bf 1} & {\bf 0}
\end{array}
\right) \ .
\end{equation}
Our parity and charge conjugation operations are defined here so to
correctly reproduce the expected transformations when restricting
to the $SU_{L}(N_f)\times SU_{R}(N_f)$ subgroup. Since ${\cal C}$
is an element of $SU(2N_f)$ global invariance automatically
guarantees the charge conjugation one.

The $\epsilon $ Lagrangian respecting global chiral rotations is:
\begin{eqnarray}
\tilde{\Gamma}_{WZ}\left[ U,\;A\right] &=&\tilde{\Gamma}_{WZ}\left[ U\right]
\,+\,C_{1}i\int_{M^{4}}{\rm Tr}\left[ A\alpha ^{3}\right]  \nonumber \\
&&-C_{2}\,\int_{M^{4}}{\rm Tr}\left[ (dAA+AdA)\alpha \right]  \nonumber \\
&&-C_{3}\int_{M^{4}}{\rm Tr}\left[ dAdUA^{T}U^{-1}-dA^{T}dU^{-1}AU \right]
\nonumber \\
&&-C_{4}\,\int_{M^{4}}{\rm Tr}[UA^{T}U^{-1}(A\alpha ^{2}+\alpha ^{2}A)]
\nonumber \\
&&+\,C_{5}\int_{M^{4}}{\rm Tr}\left[ (A\alpha )^{2}\right]
+C_{6}\,i\int_{M^{4}}{\rm Tr}\left[ A^{3}\alpha \right]  \nonumber \\
&&+C_{7}\,i\int_{M^{4}}{\rm Tr}\left[ (dAA+AdA)UA^{T}U^{-1}\right]  \nonumber
\\
&&-C_{8}i\int_{M^{4}}{\rm Tr}\left[ A\alpha AUA^{T}U^{-1}\right]
+C_{9}\,\int_{M^{4}}{\rm Tr}[A^{3}UA^{T}U^{-1}]  \nonumber \\
&&+C_{10}\int_{M^{4}}{\rm Tr}[(AUA^{T}U^{-1})^{2}]  \nonumber \\
&&+C_{11}i\int_{M^{4}}{\rm Tr}[A^{2}(\alpha UA^{T}U^{-1}-UA^{T}U^{-1}\alpha
)] \ ,
\end{eqnarray}
where $C_{i}$ are imaginary. The last term is a new term not
generated by gauging the Wess-Zumino effective action. If we
require the Lagrangian to respect the continuous enhanced global
symmetry $Sp(2N_{f})\times SU(2N_{f})$, no vector axial $\epsilon $
term survives. As for the $SU_{L}(N_{f})\times SU_{R}(N_{f})$ case
the enhanced symmetry imposes a very stringent constraint on the
$\epsilon $ terms. However, if only the discrete $Z_{2}$ symmetry
is imposed, several terms survive.

\section{Conclusions}

\label{conc}

In Ref.~\cite{ARS} it has been proposed that the physical spectrum of a
vector-like gauge field theory may exhibit an enhanced global symmetry near
a chiral/conformal phase transition. The new symmetry is related to the
possibility, supported by various investigations, that a parity-doubled
spectrum develops as the number of fermions $N_f$ is increased to a critical
value above which it is expected that the Goldstone phase turns into the
symmetric one.

The low energy spectrum consists of a set of spin-0 as well as
massive spin-1 particles. In this paper we have demonstrated that,
using an effective Lagrangian approach, parity-doubling, together
with the associated enhanced global symmetry, severely constrain
the $\epsilon $ terms of the effective Lagrangian describing the
interactions among vector, axial-vector and Goldstone bosons. In
particular, we showed that in the case of a continuous enhanced
symmetry (i.e. an extra $SU_{L}(N_{f})\times SU_{R}(N_{f})$ acting
only on spin-1 bosons), no dimension four $\epsilon $ term except
the ordinary Wess-Zumino one is allowed. However when we imposed
invariance only under a discrete symmetry, which also guarantees
parity doubling, a few dimension four $
\epsilon $ terms (specifically 3 for the $SU_{L}(N_{f})\times SU_{R}(N_{f})$
case) survive. If such theory is realized in nature the
phenomenology of the intrinsically negative parity terms is
predicted to be quite different from the one observed in ordinary
QCD. {}For example, no vector decay into three Goldstone bosons is
allowed.

It is important to note that if the vectors were introduced in the effective
Lagrangian as gauge bosons of the chiral symmetry, no enhanced global
symmetry would survive. In this case, enforcing the enhanced global symmetry
requires the absence of the Wess-Zumino action. This is at variance with the
t'Hooft global anomaly constraint. However just requiring global invariance
for the $\epsilon $ terms guarantees independence to the Wess-Zumino action
with respect to the extra global symmetries of the massive spectrum which is
physically acceptable since low energy theorems only apply to the massless
spectrum.

Often in literature, partially driven by vector meson dominance
considerations, the physical vector bosons are introduced as gauge bosons of
the flavor symmetries to reduce the number of free parameters in the
effective Lagrangian \cite{Schechteretal}. This seems to be a reasonable
approximation for QCD phenomenology. However if a near critical theory (with
$N_f$ large compared to $N$) develops a parity doubled spectrum with an
associated enhanced global symmetry the gauging procedure is no longer
acceptable, while the effective Lagrangian is still severely constrained. It
is, hence, plausible to expect that for large $N_f$ the enhanced symmetry
scenario is preferred over a gauged flavor symmetry suitable for low $N_f$.

Finally, we extended our investigation to the case when the
underlying fermions are in pseudo-real representations of the gauge
group. We built the $\epsilon$ effective Lagrangian terms for this
case, and then we studied its relations with respect to the
enhanced global symmetries. We are lead to the conclusions similar
to the $SU(N_f)\times SU(N_f)$ case.

\acknowledgments
It is a pleasure for us to thank Thomas Appelquist and Joseph Schechter for
helpful discussions, comments and careful reading of the manuscript. The
work of Z.D. and F.S. has been partially supported by the US DOE under
contract DE-FG-02-92ER-40704. The work of P.S.R.S. has been supported by the
Funda\c{c}\~{a}o de Amparo \`{a} Pesquisa do Estado de S\~{a}o Paulo
(FAPESP) under Contract No. 98/05643-5.

\appendix
\section{}
In this appendix we prove that all the dimension four (i.e.
4-derivative) terms involving the Lorentz tensor $\epsilon _{\mu
\nu \rho \sigma }$, consistent with global chiral symmetries as
well as $C$ and $P$ invariance, are the ones presented in
Eq.~(\ref{GWZ2}) and Eq.~(\ref{undici}) for an $SU(N)$ gauge
theory.

It is convenient to rewrite all of the right handed transforming
fields as left handed ones via:
\begin{equation}
\widetilde{A}_{L}=UA_{R}U^{-1}\ .
\end{equation}
Any term containing $A_R$ can always be expressed in terms of
$\widetilde{A}_L$ by appropriately inserting the identity operator
$U^{-1} U$. {}For example:
\begin{eqnarray}
dUdA_{R}U^{-1} &=&dUU^{-1}UdA_{R}U^{-1}=\alpha (d\tilde{A}
_{L}-dUA_{R}U^{-1}+UA_{R}dU^{-1})  \nonumber \\
&=&\alpha d\tilde{A}_{L}-\alpha ^{2}\tilde{A}_{L}-\alpha
\tilde{A}_{L}\alpha \ .
\end{eqnarray}
Since
\begin{equation}
\widetilde{\alpha}=U\beta U^{-1}=\alpha \ ,
\end{equation}
all the terms consist of products of $A_{L}$, $\widetilde{A}_{L}$,
$\alpha$ and the differential operator $d=dx^{\mu}\partial_{\mu}$.
In this way the study of independent global invariant terms is
greatly simplified.

It is more convenient to use $CP$ rather than $C$ together with $P$
when imposing invariance under the discrete symmetries. Using
Eq.~(\ref{parity}) and Eq.~(\ref{charge}) we have the following
$CP$ transformations
\begin{eqnarray}
&&A_{L}(\vec{x})\leftrightarrow
-A_{L}^{T}(-\vec{x})\;,\qquad\tilde{A}_{L}(\vec{x})\leftrightarrow
-\tilde{A}_{L}^{T}(-\vec{x})\;,  \nonumber \\ &&\alpha
(\vec{x})\leftrightarrow -\alpha ^{T}(-\vec{x})\;,\qquad
d\leftrightarrow d\;,\qquad \mbox{(measure)}\leftrightarrow
-\mbox{(measure)}
\ .
\end{eqnarray}
For reader's convenience, we provide the following identity:
\begin{equation}
{\rm Tr}\left[B_{1}^{T}B_{2}^{T}\cdots
B_{n}^{T}\right]=(-)^{\frac{\left[
\left( \sum P_{i}\right) ^{2}-\sum P_{i}^{2}\right] }{2}}{\rm
Tr}\left[B_{n}B_{n-1}\cdots B_{1} \right] \ ,
\end{equation}
where $B_{i}$ is a $P_{i}$ form.

We also need the following identities, which are very useful to
derive the gauged version of the Wess-Zumino term
\cite{Schechteretal}
\begin{equation}
\alpha ^{2n}=d\alpha ^{2n-1}\ ,\qquad \beta ^{2n}=-d\beta ^{2n-1}\ .
\label{iter}
\end{equation}

Since $\alpha $ and $A_{L}$ are algebra valued 1-forms,
\begin{equation}
{\rm Tr}\left[ A_{L}\right] ={\rm Tr}\left[ dA_{L}\right] ={\rm Tr}\left[
\alpha \right] =0\ ,
\label{1form}
\end{equation}
and
\begin{equation}
{\rm Tr}\left[ A_{L}^{2}\right] ={\rm Tr}\left[ \alpha ^{2}\right]
=0\ ,
\label{2form}
\end{equation}
while using Eq.~(\ref{iter}) we have
\begin{equation}
{\rm Tr}\left[ d\alpha \right] =0 \ .
\end{equation}
Likewise for the $\widetilde{A}_L$ field.

We are now equipped to show that non single (dimension four) traces
containing $\epsilon $ terms are absent. Indeed, given
Eq.~(\ref{1form}) and Eq.~(\ref{2form}), we only need to restrict
our analysis to any double product of the following 2-form traces:
\begin{equation}
{\rm Tr}\left[ \alpha A_{L}\right] \ ,\quad {\rm Tr}\left[ \alpha
\widetilde{A}_{L}\right] \ ,\quad {\rm Tr}\left[
{A_{L}}\widetilde{A}_{L}\right] \ .
\label{st}
\end{equation}
However any double trace term constructed as product of the single
traces presented in Eq.~(\ref{st}) is $CP$ odd and hence is not
allowed in the effective Lagrangian. So we are left to investigate
the dimension four single trace terms.

\[
.
\centerline{
\begin{tabular}{|l||llll||l|l|}
\hline
& $n_{1}$ & $n_{2}$ & $n_{3}$ & $n_{4}$ & Tr[ Term  ]& New ? \\
\hline $1$ & $4$ & $0$ & $0$ & $0$ & $%
A_{L}^{4}=A_{L}^{3}A_{L}=-A_{L}A_{L}^{3}=-A_{L}^{4}=0$ (also $CP$
odd) & $No$
\\ \hline
$2$ & $3$ & $1$ & $0$ & $0$ & $A_{L}^{3}\tilde{A}_{L}$ & $Yes$ \\
\hline $3$ & $3$ & $0$ & $1$ & $0$ & $A_{L}^{3}\alpha $ & $Yes$ \\
\hline $4$ & $3$ & $0$ & $0$ & $1$ & $A_{L}^{2}dA_{L}=d(\frac{1}{3}
A_{L}^{3})\rightarrow 0$ \ (also $CP$ odd) & $No$ \\
\hline $5$ &
$2$ & $2$ & $0$ & $0$ & $A_{L}^{2}\tilde{A}_{L}^{2}\stackrel{CP}{
\rightarrow }-(A_{L}^{T})^{2}(\tilde{A}_{L}^{T})^{2}=-\tilde{A}
_{L}^{2}A_{L}^{2}=-A_{L}^{2}\tilde{A}_{L}^{2}\rightarrow 0$ & $No$ \\
\cline{1-1}\cline{6-7}
$6$ &  &  &  &  & $A_{L}\tilde{A}_{L}A_{L}\tilde{A}_{L}$ & $Yes$ \\
\hline $7$ & $2$ & $1$ & $1$ & $0$ & $A_{L}^{2}\tilde{A}_{L}\alpha
$ & $Yes$ \\
\cline{1-1}\cline{6-7}
$8$ &  &  &  &  & $A_{L}^{2}\alpha
\tilde{A}_{L}\stackrel{CP}{\rightarrow }
-(A_{L}^{T})^{2}\alpha ^{T}\tilde{A}_{L}^{T}=-\tilde{A}_{L}\alpha
A_{L}^{2}=-A_{L}^{2}\tilde{A}_{L}\alpha $ & $No$ \\ \cline{1-1}\cline{6-7}
$9$ &  &  &  &  & $A_{L}\tilde{A}_{L}A_{L}\alpha $ & $Yes$ \\ \hline
$10$ & $2$ & $1$ & $0$ & $1$ & $dA_{L}A_{L}\tilde{A}_{L}$ & $Yes$ \\
\cline{1-1}\cline{6-7}
$11$ &  &  &  &  &
$A_{L}dA_{L}\tilde{A}_{L}\stackrel{CP}{\rightarrow }
A_{L}^{T}dA_{L}^{T}\tilde{A}_{L}^{T}=-\tilde{A}_{L}dA_{L}A_{L}=dA_{L}A_{L}
\tilde{A}_{L}$ & $No$ \\ \cline{1-1}\cline{6-7}
$12$ &  &  &  &  &
$A_{L}^{2}d\tilde{A}_{L}\stackrel{CP}{\rightarrow }
(A_{L}^{T})^{2}d\tilde{A}_{L}^{T}=-d\tilde{A}_{L}A_{L}^{2}=-A_{L}^{2}d
\tilde{A}_{L}\rightarrow 0$ & $No$ \\ \hline
$13$ & $2$ & $0$ & $2$ & $0$ & $A_{L}^{2}\alpha ^{2}\stackrel{CP}{
\rightarrow }-(A_{L}^{T})^{2}(\alpha ^{T})^{2}=-\alpha
^{2}A_{L}^{2}=-A_{L}^{2}\alpha ^{2}\rightarrow 0$ & $No$ \\
\cline{1-1}\cline{6-7}
$14$ &  &  &  &  & $(A_{L}\alpha )^{2}$ & $Yes$ \\ \hline
$15$ & $2$ & $0$ & $1$ & $1$ & $dA_{L}A_{L}\alpha $ & $Yes$ \\
\cline{1-1}\cline{6-7}
$16$ &  &  &  &  & $A_{L}dA_{L}\alpha \stackrel{CP}{\rightarrow }
A_{L}^{T}dA_{L}^{T}\alpha ^{T}=-\alpha
dA_{L}A_{L}=dA_{L}A_{L}\alpha $ & $No$
\\ \hline
$17$ & $2$ & $0$ & $0$ & $2$ & $(dA_{L})^{2}=d(A_{L}dA_{L})\rightarrow 0$
(also $CP$ odd) & $No$ \\ \hline
$18$ & $1$ & $1$ & $2$ & $0$ & $A_{L}\tilde{A}_{L}\alpha ^{2}$ & $Yes$ \\
\cline{1-1}\cline{6-7}
$19$ &  &  &  &  & $A_{L}\alpha
^{2}\tilde{A}_{L}\stackrel{CP}{\rightarrow}
-A_{L}^{T}(\alpha ^{T})^{2}\tilde{A}_{L}^{T}=-\tilde{A}_{L}\alpha
^{2}A_{L}=A_{L}\tilde{A}_{L}\alpha ^{2}$ & $No$ \\ \cline{1-1}\cline{6-7}
$20$ &  &  &  &  & $A_{L}\alpha \tilde{A}_{L}\alpha \stackrel{P}{\rightarrow
}-A_{L}\alpha \tilde{A}_{L}\alpha \rightarrow 0$ ($CP$ even) & $No$ \\ \hline
$21$ & $1$ & $1$ & $1$ & $1$ & $dA_{L}\alpha \tilde{A}_{L}$ & $Yes$ \\
\cline{1-1}\cline{6-7}
$22$ &  &  &  &  & $dA_{L}\tilde{A}_{L}\alpha
\stackrel{CP}{\rightarrow} dA_{L}^{T}\tilde{A}_{L}^{T}\alpha
^{T}=-\alpha \tilde{A}_{L}dA_{L}=-dA_{L}
\alpha \tilde{A}_{L}$ & $No$ \\ \cline{1-1}\cline{6-7}
$23$ &  &  &  &  & $A_{L}d\tilde{A}_{L}\alpha =-d(A_{L}\tilde{A}_{L}\alpha
)+dA_{L}\tilde{A}_{L}\alpha +A_{L}\tilde{A}_{L}\alpha ^{2}$ & $No$ \\
\cline{1-1}\cline{6-7}
$24$ &  &  &  &  & $A_{L}\alpha d\tilde{A}_{L}=d(A_{L}\alpha
\tilde{A}
_{L})-dA_{L}\alpha \tilde{A}_{L}+A_{L}\alpha ^{2}\tilde{A}_{L}$ & $No$ \\
\hline
$25$ & $1$ & $1$ & $0$ & $2$ &
$dA_{L}d\tilde{A}_{L}=d(A_{L}d\tilde{A}
_{L})\rightarrow 0$ (also $CP$ odd) & $No$ \\ \hline
$26$ & $1$ & $0$ & $3$ & $0$ & $A_{L}\alpha ^{3}$ & $Yes$ \\ \hline
$27$ & $1$ & $0$ & $2$ & $1$ & $dA_{L}\alpha ^{2}=d(A_{L}\alpha
^{2})\rightarrow 0$ (also $CP$ odd) & $No$ \\ \hline
$28$ & $0$ & $0$ & $4$ & $0$ & $\alpha ^{4}=0$ (See $4000$ case,
also $CP$ odd) & $No$ \\
\hline
\end{tabular}}
\]
\begin{table}[h]
\label{table}
\caption{List of the allowed $\epsilon$ single trace, 4-derivative terms.
The {\it Yes} terms are considered to be independent while the
terms labeled as {\it No} are either non independent or forbidden
as indicated. Each single term is understood to be under the Tr
sign.}
\end{table}

By enumerating all of the single trace, dimension four (i.e.
4-derivative), terms involving the Lorentz tensor $\epsilon
_{\mu \nu \rho\sigma }$, consistent with chiral symmetries as well as $CP$ and $P$
invariance, we are ready to demonstrate that the terms presented in
Eq.~(\ref{GWZ2}) together with Eq.~(\ref{undici}) exhaust all the
possibilities for an $SU(N)$ gauge theory.

As mentioned before, any 4-form term can be written as a
combination of $n_{1}$ $A_{L}$, $n_{2}$ $\widetilde{A}_{L}$,
$n_{3}$ $\alpha $ and $n_{4}$ $d$ with $\sum n_{i}=4$. We can
restrict the count of independent terms to those with $n_{1}\geq
n_{2}$ since the ones with $n_{1} < n_{2}$ are related by a $P$
transformation. {}For non vanishing terms we have $n_{4}\leq 2$
since $d^{2}=0$. Besides, $d$ should not act on $\alpha$,
otherwise, via Eq.~(\ref{iter}), we introduce double counting. All
of the possible combinations are summarized in Table I. We trace
over each term while we omit an explicit ${\rm Tr}$ symbol.

The Table I proves that we have only $11$ independent terms.
Finally by expanding the latter and providing their $P$ and $C$
partners, we deduce Eq.~(\ref{GWZ2}) and Eq.~(\ref{undici}) as the
most general dimension 4 effective Lagrangian involving the Lorentz
tensor $\epsilon_{\mu \nu\rho \sigma }$.

\end{document}